\documentclass{JHEP3}
\usepackage{amsmath} 
\usepackage{amssymb} 
\usepackage{amscd}
\newcommand{\scomma}{\, ,\,}
\newtheorem{theorem}{Theorem}
\newtheorem{corollary}{Corollary}
\newenvironment{proof}[1][Proof]{\begin{trivlist}
\item[\hskip \labelsep {\bfseries #1}]}{\end{trivlist}}
\title{Remarks on the Formulation of Quantum Mechanics on Noncommutative Phase Spaces}
\author{B. Muthukumar \\Saha Institute of Nuclear Physics,\\ 1/AF, Bidhan nagar, Kolkata-700 064, India.\\ E-mail: \email{b.muthukumar@saha.ac.in}}
\abstract{
We consider the probabilistic description of nonrelativistic, spinless one-particle classical mechanics, and immerse the particle in a deformed noncommutative phase space in which position coordinates do not commute among themselves and also with canonically conjugate momenta. With a postulated normalized distribution function in the quantum domain, the square of the Dirac delta density distribution in the classical case is properly realised in noncommutative phase space and it serves as the quantum condition. With only these inputs, we pull out the entire formalisms of noncommutative quantum mechanics in phase space and in Hilbert space, and elegantly establish the link between classical and quantum formalisms and between Hilbert space and phase space formalisms of noncommutative quantum mechanics. Also, we show that the distribution function in this case possesses 'twisted' Galilean symmetry. 
}
\preprint{\hepth{0609117}}
\begin{document}
%%%%%%%%%%%%%%%%%%%%%%%%%%%%%%%%%%%%%%%%%%%%%%%%%%%%%%%
\section{Introduction}
%%%%%%%%%%%%%%%%%%%%%%%%%%%%%%%%%%%%%%%%%%%%%%%%%%%%%%%
%
Several arguments are presently used in the literature to motivate a noncommutative (NC) structure of spacetime coordinates at very short distances, especially it has been shown in \cite{Seiberg} that quantum field theories in NC spacetime naturally arise as a decoupling limit of worldvolume dynamics of D-brane in a constant NS-NS two form background (for reviews, see \cite{Douglas}). As it might be easier to understand the effects of spatial noncommutativity in simpler setups, nonrelativistic sector of NC field theories and their one particle sectors in the free field or weak coupling case--  the noncommutative quantum mechanics (NCQM)--have been extensively investigated \cite{Gomis,Chaichian}. 

Recently, using a modified form of Wigner distribution function (WDF) the authors of \cite{Jing} attempted to give a phase space formulation of NCQM and showed that the modified WDF satisfies a generalized $*$-genvalue equation for Hamiltonian, given that the corresponding Hamiltonian operator satisfies the eigenvalue equation in Hilbert space. We extend their work in several ways. Our starting point is \cite{Baker} where, using the concept of Moyal's sine bracket and a cosine bracket and a quantum condition, Baker attempted to establish a physically manageable, postulational formulation of QM in phase space. Baker also showed how the WDF of QM naturally arises from the postulated quantum condition. This program  was further pursued in \cite{Fairlie,Curtright,Dias1,Dias2} and the mathematical tools to establish its link with Hilbert space formalism were further developed. However, much remains to be done to unify the different strands and to concretely establish the link on the one hand between the mathematical formalism of classical mechanics and that of quantum mechanics, and on the other  hand between the phase space and Hilbert space formalisms of QM. We address these issues in the context of NCQM. We start with the probabilistic description of one-particle classical mechanics in phase space, using Dirac delta distributions. In the quantum domain we introduce the noncommutativity through the concept of deformed $*$-product. We carry forward the concept of normalized distribution function $f(x,p)$ to the quantum domain and use the quantum condition $f*f = \kappa f$, where $\kappa$ is some constant to be determined. It turns out that the bare NC geometrical structure of phase space and the  normalized phase space distribution function with its quantum condition and with the 'classical' definition of expectation value for a physical observable are enough to extract (i) the whole formalism of NCQM of nonrelativistic spinless particles in phase space with its eigenvalue equation and the Moyal dynamical equation and (ii) the whole formalism of NCQM of the same in Hilbert space including the concept of wave function, the operator algebra, Schr\"{o}dinger equation and the Heisenberg equation.

The paper is organized as follows. To motivate our use of a modified form of Baker's quantum condition, we discuss briefly in Section 2 the one-particle classical mechanics in phase space in a manner that brings to the fore the problem-setting and the physical concepts to be introduced. We use product of Dirac delta functions as a probability distribution to describe the distribution of a particle in phase space. In Section 3, which is divided into four subsections, we first introduce general ideas about NC geometry and deformation of ordinary products of two functions, and briefly review how NC spaces are constructed in the deformation context. Given the deformed NC geometric (algebraic) structure, we postulate the normalized distribution function $f(x,p)$ and the above mentioned quantum condition for it in the quantum domain and point out how the quantum condition may be inferred from a classical concept. From the quantum condition we construct a noncommutative Fredholdm integral equation of second kind and show that noncommutative Hilbert-Schmidt theory of integral equations demands that $\kappa$ be directly related to the parameter of deformation in $x-p$ planes and that $1/\kappa$ be integrally quantized. We identify that integer with the number of non-interfering equally probable mixed states. For a pure state the solution of that integral equation leads to an explicit form for probability distribution in terms of what we would be later identifying as the quantum mechanical wave function. This expression turns out to be that of Wigner's quasi-probability distribution function (WDF) in NC space obtained in \cite{Jing} through a modification of Weyl-Wigner correspondence. We write this WDF in a new form using the deformed products. It turns out that this new form holds the key to the elegant establishment of the link between phase space and Hilbert space formalisms of QM (NCQM), or more precisely, the link between the dynamical and eigenvalue equations in these two formalisms. This new form also motivates us to prove a theorem that establishes the link between the star product and operator product. This we do in Section 4. We examine the possible small variations of the distribution function and of the wave function that would preserve the normalization and quantum conditions. We then give a variational method to derive  integral forms of stationary state eigenvalue equations for physical observables in the two formalisms and  clearly identify the point of separation of these formalisms. Section 5 discusses how those integral forms of eigenvalue equations are converted into the conventional ones. Section 6 deals with the dynamical equations and elegantly establishes the link between the two formalisms. In Section 7, we show that the NC WDF possesses not the Galilean symmetry but what is known as 'twisted' Galilean symmetry \cite{Chakraborty}. We end with concluding remarks in Section 8.
%
%%%%%%%%%%%%%%%%%%%%%%%%%%%%%%%%%%%%%%%%%%%%%%%%%%%%%%%%%%%%%
\section{Classical Domain}
%%%%%%%%%%%%%%%%%%%%%%%%%%%%%%%%%%%%%%%%%%%%%%%%%%%%%%%%%%%%%
%
The phase space of classical mechanics represents the configuration space denoted by the set  of points $ (x, p) $ of positions and momenta. A particle in phase space is represented as a perfectly localized object at a 'point' $(x_0(t),p_0(t))$ at time $t$, and its probability density distribution $ f(x,p,t) $ in a $2n$ dimensional phase space may be represented using Dirac delta distributions:
\begin{eqnarray}
     f(x,p,t) = \delta^{(n)}(x-x_0(t)) \delta^{(n)}(p-p_0(t)) \label{classicalf}
\end{eqnarray} 
Physical observables are real functions $ A(x,p) $ on phase space. The phase space average of a physical quantity $ A(x,p) $ is, then, given by 
\begin{eqnarray}
     \left< A \right> &=& \frac{\int\; d^n x \; d^n p\; A(x,p) f(x,p,t) }{\int \; d^n x \; d^n p\; f(x,p,t)} \label{expect} \\
     &=& A(x_0(t),p_0(t))  \label{expect0} %\label{expect1}
\end{eqnarray} 
If we assume that the Dirac delta densities are allowed density distributions in phase space, then the evolution of probability density must follow the Liouville equation:
\begin{eqnarray}
     \frac{\partial f}{\partial t} &=&\delta^{ij} \left(\frac{\partial H}{\partial x^i} \frac{\partial f}{\partial p^j}-\frac{\partial H}{\partial p^j} \frac{\partial f}{\partial x^i}\right) \nonumber \\
     &=& \left\{ H, f \right\}, \label{Liouville}
\end{eqnarray} 
where $\{ \}$ denotes the classical Poisson bracket and $ H $ the Hamiltonian. Using  (\ref{expect}) and (\ref{Liouville}), we may arrive at the evolution of the expectation values of positions and momenta
\begin{eqnarray}
     \frac{d x^i_0(t)}{dt}  = \frac{\partial H}{\partial p^i}; \qquad \frac{d p^i_0(t)}{dt} = - \frac{\partial H}{\partial x^i}, \nonumber
\end{eqnarray} 
which are the Hamilton equations that determine the particle trajectory. 
%
%%%%%%%%%%%%%%%%%%%%%%%%%%%%%%%%%%%%%%%%%%%%%%%%%%%%%%%%%%%%%
\section{Quantum Domain}
%%%%%%%%%%%%%%%%%%%%%%%%%%%%%%%%%%%%%%%%%%%%%%%%%%%%%%%%%%%%%
%
%%%%%%%%%%%%%%%%%%%%%%%%%%%%%%%%%%%%%%%%%%%%%%%%%%%%%%%%%%%%%
\subsection{Deformed Products}
%%%%%%%%%%%%%%%%%%%%%%%%%%%%%%%%%%%%%%%%%%%%%%%%%%%%%%%%%%%%%
%
In \cite{Baker} a different road to quantization in phase space was outlined. The basic idea in that work is to carry forward the definition of expectation value (\ref{expect}) to the quantum domain, replace the composition of functions by Moyal's sine bracket and a cosine bracket, and postulate a quantum condition satisfied by normalized $f$. The results obtained there are very interesting when they are reread in the light of the ideas of deformation theory and of modern NC geometry. Therefore, before discussing our objectives stated in the Introduction, we introduce the general ideas about deformation theory and its connection with NC geometry. 

Classical geometry is based on the concept of set of points which we call 'space'. In many cases, however, the geometrical objects such as curves, surfaces etc., are better studied not as sets of points but by analysing the properties of certain commutative algebras of functions on such sets or spaces \cite{Landi}. In fact, it turns out that there is a duality between certain categories of geometric spaces and categories of commutative algebras of functions on those spaces \cite{Khalkhali}, and therefore studying those algebras is the same as studying  the spaces themselves. 

The starting point in the analysis of NC geometric spaces is to drop the commutativity in their representing algebras, and in many cases the individual functions themselves may be defined on ordinary 'commutative' spaces. In many cases, the sets of points associated with these algebras do not even exist 'concretely'. However, this is irrelevant for the purpose of studying those spaces as all the needed informations are encoded in the NC algebras. In the case of what is termed the NC geometry 'in the small' \cite{Ginzburg}, the noncommutativity in the algebras is 'introduced' through the concept of deformation. 

Consider a $D$-dimensional Euclidean ${\mathbb R}^{D}$ manifold ${\mathcal M}$ parameterized by $D$ coordinates $X^{\alpha}$, and endowed with a ``Poisson structure'' $\Sigma$ ---a skew-symmetric bidifferential operation of Poisson bracket on functions on that manifold. Let $\nabla_{\mu}$ be the operator of covariant differentiation defined in terms of the Poisson connection on that manifold ---a connection without torsion and curvature such that $\nabla_{\mu} \Sigma = 0$. We define the ordinary commutative product of two functions $A$ and $B$ as $AB$ and use $*$ to distinguish the NC product $A*B$. The later product is defined as a ``deformation'' of commutative product by a smooth function
\begin{align}
     g(z) = \sum_{r=0}^{\infty} a_r \frac{z^r}{r!}, \qquad a_0= a_1=1, \label{deformfun}
\end{align} 
and it is expressed as
\begin{align}
     A*B = \sum_{r=0}^{\infty} \left(\frac{i}{2}\right)^r \frac{a_r}{r!} \Sigma^r(A,B) \label{deform*}
\end{align} 
where
\begin{align}
     \Sigma^r(A,B) = \Sigma^{\alpha_1 \beta_1} \ldots \Sigma^{\alpha_r \beta_r} \left\{\nabla_{\alpha_1}\ldots \nabla_{\alpha_r} A \right\} \left\{ \nabla_{\beta_1} \ldots \nabla_{\beta_r}B \right\}  \nonumber
\end{align} 
with $ \Sigma^{\alpha \beta} = -\Sigma^{\beta \alpha}$. It turns out that when $\Sigma^{\alpha \beta}$ is a constant matrix, the only function (\ref{deformfun}) that can preserve the associativity of $*$-product is the exponential function \cite{Bayen1}, i.e., 
\begin{align}
     (A*B)*C  =  A*(B*C) \label{asso}
\end{align}
iff $ a_r = 1 \;\forall \;\,r$. The reason for the insertion of the factor $(i/2)^r$ in (\ref{deform*}) is that the generators $X^{\alpha} \in {\mathbb R}$ of the deformed algebra satisfy a familiar commutation relation $ X^{\alpha}~ *~X^{\beta}-X^{\beta}~*~X^{\alpha} = i \Sigma^{\alpha\beta}$. Since we want to study the particle mechanics in $2n$ dimensional phase space ${\mathbb R}^{n}\times {\mathbb R}^{n}$ in the quantum domain where not only that $x^i$ do not commute with $p^i$ but also that $x^i$ do not commute with $x^j$ for $i\neq j$, we take $D=2n$, $ \{ X^{\alpha}\} = \{x^i, p^j | x^i, p^j \in {\mathbb R}\}$ and 
\begin{eqnarray}
     \Sigma^{\alpha\beta} = \left(\begin{array}{cc}
     \theta^{ij} &  \hbar \delta^k_l \\
     - \hbar \delta^k_l & \mathbf{0}
\end{array} \right),\nonumber
\end{eqnarray} 
where $\theta^{ij} (= -\theta^{ji})$ is constant and real, and it characterizes deformation along  spatial directions; $\hbar$ is the parameter of deformation in $x$-$p$ planes (we will  initially take these two deformations to be independent and later deduce their hierarchy). Then the commutation relations between $x^i$ and $p^j$ become 
\begin{align}
     &\left[x_i \scomma x_j\right] = x_i * x_j - x_j * x_i = i \theta_{ij}; \label{canonical} \\
     &\left[x_i \scomma p_j\right] = i \hbar \delta_{ij}; \qquad
     \left[p_i\scomma p_j\right] = 0, \label{werner}
\end{align} 
and the star product takes the explicit form:
\begin{align}
      A(x,p) * B(x,p) = \left. e^{\left\{\frac{i\hbar}{2}\left(\frac{\partial}{\partial x_i}\frac{\partial}{\partial p'_i}-\frac{\partial}{\partial x'_i}\frac{\partial}{\partial p_i}\right)+\frac{i\theta^{ij}}{2}\left(\frac{\partial}{\partial x_i}\frac{\partial}{\partial x'_j}\right)\right\}} A(x,p) B(x',p') \right|_{\substack{x'=x\\p'=p}} \label{prep}
\end{align} 
and it has the Fourier space representation: 
\begin{align}
     A(x,p)*B(x,p) 
     &= \frac{1}{(2\pi)^{4n}}\int d^n k \, d^n l \,d^n k' \, d^n l' \, \tilde{A}(k,l) \tilde{B}(k',l') \times \nonumber \\ 
     & \phantom{\int d^n k \, d^n l} \times e^{-\frac{i\hbar}{2} (k_i l'_i - k'_i l_i)}   e^{-\frac{i}{2} \theta^{ij} k_i k'_j} e^{i(k_i + k'_i) x^i + i (l_i +l'_i)p^i}. \label{krep}
\end{align} 

The above expression yields ordinary commutative product between two proper functions under integration over full phase space:
\begin{align}
    \int d^n x d^n p A(x,p) * B(x,p) =  \int d^n x d^n p A(x,p) B(x,p) 
    \label{star2ordi}
\end{align}

Owing to the above property, the Moyal bracket of two proper functions vanishes under integration
\begin{align}
    \int d^n x d^n p \left[A \scomma B \right] = 0. \label{zeroMB}
\end{align} 

From the relation (\ref{star2ordi}), we may deduce the cyclicity property under integration:
\begin{align}
    \int d^n x d^n p \, A(x,p) * B(x,p) * C(x,p) = \int d^n x d^n p \, C(x,p) * A(x,p)* B(x,p). \label{cyclic}
\end{align} 

Since the phase in the exponential factor in (\ref{prep}) is antisymmetric under the exchange of primed and unprimed $x$ and $p$, the complex conjugation of the star product becomes
\begin{equation}
      (A(x,p) * B(x,p))^* = B^*(x,p) * A^*(x,p), \label{cc}
\end{equation} 
where the $*$ as a superscript denotes the complex conjugation. Therefore, $ A*A$ is real if $A$ is real. But in general $A*B$ is complex even if $A$ and $B$ are real. In the following, we will also be dealing with $*_{\hbar}$-product and $*_{\theta}$-product which are respectively obtained by putting $\theta = 0$ and $\hbar=0$ in (\ref{prep}) or in (\ref{krep}). These products also have the above properties (\ref{asso}), (\ref{star2ordi}), (\ref{zeroMB}), (\ref{cyclic}) and (\ref{cc}). 
%
%%%%%%%%%%%%%%%%%%%%%%%%%%%%%%%%%%%%%%%%%%%%%%%%%%%%%%%
\subsection{Quantum Condition}
%%%%%%%%%%%%%%%%%%%%%%%%%%%%%%%%%%%%%%%%%%%%%%%%%%%%%%%
%
In the foregoing analysis, the bare geometrical (algebraic) structure was postulated. To do physics in such a NC phase space, following \cite{Baker} we postulate the existence of a distribution function and use the same definition of expectation value (\ref{expect}) in the quantum domain. Then we ask what is the 'expectation value' of the density distribution itself, or rather, what is the square of (Dirac delta) density distribution? The answer to this mathematically discomforting question seems to possess much deeper meaning once we replace the composition of two functions through ordinary product in phase space by the composition through star product.  In \cite{Baker}, the starting point was the postulation of a normalized real distribution function
\begin{align}
     \int d^n x \; d^n p \; f(x,p,t) = 1, \label{normal}
\end{align}  
with the quantum condition:
\begin{align}
     f *_{\hbar} f = (2\pi\hbar)^{-n} f, \label{Bakerqcond}
\end{align} 
which, in the limit $\hbar \to 0 $, would reproduce the corresponding relation in classical phase space. We will show, however, that the strict relationship (\ref{Bakerqcond}) may be relaxed and made $ f *_{\hbar} f \propto f $, and the inverse of proportionality constant is integrally 'quantized'. We identify that integer with the number of mixed, equally probable and  noninterfering states. To make sure of the generality of the procedure we adopt, we work with the $*$-product instead of $*_{\hbar}$-product and postulate the quantum condition:
\begin{align}
     f * f = \kappa f \label{*qcond}, 
\end{align} 
where $\kappa$ is the proportionality constant to be determined. (Incidentally, it is easy to show using (\ref{krep}) that the $*_{\hbar}$-composition of two $f$'s represented by (\ref{classicalf}) gives only the constant $(\pi\hbar)^{-2n}$. Therefore, it does not correctly reproduce the result $ f^2 = (1/0^+) f $ in the limit $\hbar \to 0 $. The reason for this anomaly is that in NC phase space the concept of point does not exist which calls for a modification of $f$ itself in NC phase space.) 

We stress that the bare NC geometrical structure and the normalized distribution function with its quantum condition and with the definition of expectation value (\ref{expect}) are enough to extract the whole of quantum mechanical formalisms of nonrelativistic spinless particles in phase space and in Hilbert space, and to establish the link between them. 
%
%%%%%%%%%%%%%%%%%%%%%%%%%%%%%%%%%%%%%%%%%%%%%%%%%%%%%%%%%%%%
\subsection{Baker's Construction of Integral Equation}
%%%%%%%%%%%%%%%%%%%%%%%%%%%%%%%%%%%%%%%%%%%%%%%%%%%%%%%%%%%%
%
Next, we examine the consequences of (\ref{*qcond}) to an explicit form of $f$. For this purpose we want to construct an integral equation out of (\ref{*qcond}), following the method of \cite{Baker}. First we generalize Baker's method to the star product of two functions of the form 
\begin{align}
A_3(x,p) = A_1(x,p) * A_2(x,p) \label{A1*A2}
\end{align} 
in the NC phase space characterized by (\ref{canonical}) and (\ref{werner}), and then analyse the special case (\ref{*qcond}). Later this generalization will also be useful for the derivation of eigenvalue equation which was not dealt with in \cite{Baker}. 

In Fourier space the star product of two functions has the following expression.
%
%\begin{align}
%     A_3(x,p) &= A_1(x,p) * A_2(x,p) \\
%     &= \frac{1}{(2\pi)^{4n}}\int d^n k \, d^n l \,d^n k' \, d^n l' \, \tilde{A}_1(k,l) \tilde{A}_2(k',l') \times \nonumber \\ 
%     &\phantom{\int d^n k \, d^n l} \times e^{-\frac{i\hbar}{2} (k_i l'_i - k'_i l_i)}   e^{-\frac{i}{2} \theta^{ij} k_i k'_j} e^{i(k_i + k'_i) x^i + i (l_i +l'_i)p^i} %\label{A1*A2}
%\end{align} 
%
Let $ \bar{A}_a(x,l)$, where $a = 1,2,3$, be the Fourier transforms of the momentum dependencies of $A_a(x,p)$. Then,
\begin{align}
     \bar{A}_a(x,w) &= \int d^n p e^{-iw_i p^i} A_a(x,p); \label{Abar}\\
     \tilde{A}_a(k,l) &= \int d^n x e^{-ik_i x^i} \bar{A}_a(x,l) \label{Atilde}.
\end{align} 
If we take the Fourier transform of momentum dependencies on both sides of (\ref{A1*A2}), we get 
\begin{align}
     \bar{A}_3(x,w) &= \frac{1}{(2\pi)^{4n}}\int d^n k \, d^n l \,d^n k' \, d^n l' \, d^n p \, d^n y \, d^n z \, \bar{A}_1(y,l) \bar{A}_2(z,l') \times \nonumber \\
     & \phantom{\frac{1}{(2\pi)^{4n}}\int}\times e^{i(l_i + l'_i - w_i ) p^i } e^{i(x^i + \frac{\theta^{ij}}{2} k_j - z^i + \frac{\hbar}{2} l^i) k'_i} e^{i(x^i - y^i -\frac{\hbar}{2} l'_i) k_i},\nonumber
\end{align} 
where Fourier space representation of star product (\ref{krep}) has been made use of. Straightforward integration over $ k', p, l$ and $z$ gives
\begin{align}
     \bar{A}_3(x,w) &= \frac{1}{(2\pi)^{2n}}\int d^n k \, d^n l' \, d^n y \, 
      \, e^{i(x^i - y^i -\frac{\hbar}{2} l'_i) k_i} \times\nonumber \\
     & \phantom{\frac{1}{(2\pi)^{4n}}\int}\times \bar{A}_1(y,\, w-l') \bar{A}_2(x^i + \frac{\theta^{ij}}{2} k_j - \frac{\hbar}{2} (l'^i-w^i), \, l')  \nonumber \\
     &= \frac{1}{(2\pi)^{2n}}\int d^n k \, d^n l' \, d^n y \, e^{i(x^i - y^i -\frac{\hbar}{2} l'_i) k_i}
      \times \nonumber \\
     & \phantom{\frac{1}{(2\pi)^{4n}}\int}\times \bar{A}_1(y,\, w-l') \bar{A}_2(x^i -i \frac{\theta^{ij}}{2} \frac{\overset{\leftarrow}{\partial}}{\partial x^j} - \frac{\hbar}{2} (l'^i-w^i), \, l') , \nonumber
\end{align} 
where in the last step we have replaced $k_j$ in the argument of $\bar{A}_2 $ by $-i \frac{\overset{\leftarrow}{\partial}}{\partial x^j} $ which acts on $ e^{ik_ix^i}$. Performing the integration over $k$ and $y$, we get
\begin{align}
     \bar{A}_3(x,w) 
     &= \frac{1}{(2\pi)^{n}}\int d^n l' \, 
      \bar{A}_2(x^i -i \frac{\theta^{ij}}{2} \frac{\vec{\partial}}{\partial x^j} - \frac{\hbar}{2}( l'^i-w^i), \, l'^i) \bar{A}_1(x-\frac{\hbar}{2}l',\, w-l') \nonumber \\
     &= \frac{1}{(2\pi)^{n}}\int d^n l' \, \bar{A}_1(x-\frac{\hbar}{2}l',\, w-l') *_{\theta} \bar{A}_2(x - \frac{\hbar}{2}( l'-w), \, l') \label{abar0}
\end{align} 
The change of integration variable $ l=-\hbar l' + \frac{\hbar}{2}w $ in (\ref{abar0}) returns
\begin{eqnarray}
\begin{aligned}
     \bar{A}_3(x,w) 
     &= \frac{1}{(2\pi\hbar)^{n}}\int d^n l \, \bar{A}_1\left(x +\frac{1}{2}(l-\frac{\hbar }{2}w),\, \frac{1}{\hbar}(l+\frac{\hbar}{2}w)\right) *_{\theta} \\ & \phantom{\frac{1}{(2\pi\hbar)^{n}}\int d^n l \,} *_{\theta} \bar{A}_2\left(x+\frac{1}{2}(l+\frac{\hbar}{2}w),\, -\frac{1}{\hbar}(l-\frac{\hbar}{2}w)\right) \label{Abar1}
\end{aligned}
\end{eqnarray} 
Next, consider the case of (\ref{Abar1}) in which $ w=0$:
\begin{align}
     \bar{A}_3(x,0) 
     &= \frac{1}{(2\pi\hbar)^{n}}\int d^n l \, \bar{A}_1\left(x+\frac{1}{2}l,\, \frac{l}{\hbar}\right) *_{\theta} \bar{A}_2\left(x+ \frac{1}{2}l,\, -\frac{l}{\hbar}\right), \label{Abar2}
\end{align} 
and let $\bar{A}_1\left(x+\frac{1}{2}l,\, \frac{l}{\hbar}\right) = \bar{B}_1(x,l+x)$. Then, 
\begin{align}
     \bar{A}_1(x,l) = \bar{B}_1(x-\frac{\hbar}{2}l, \, x+\frac{\hbar}{2}l). \label{A1bar}
\end{align} 
For real
$A(x,p)$, Eq.(\ref{Abar}) tells that $ \bar{A}_1(x,l)= \bar{A}_1^*(x,-l)$. Therefore, $\bar{B}_1^*(x-\frac{\hbar}{2}l, x+\frac{\hbar}{2}l) = \bar{B}_1(x+\frac{\hbar}{2}l, x-\frac{\hbar}{2}l)$ (or) $\bar{B}_1^*(x,l)=\bar{B}_1(l,x)$. Similarly we define $\bar{A}_2\left(x+ \frac{1}{2}l,\, \frac{1}{\hbar}l\right) = \bar{B}_2(x,l+x)$ and we get $\bar{B}_2^*(x,l)=\bar{B}_2(l,x)$ for real $ B(x,p)$. Further, if it is that 
\begin{eqnarray}
     A_3(x,p) = \alpha A_2(x,p), \nonumber
\end{eqnarray} 
for some constant $\alpha$, then (\ref{A1*A2}) becomes
\begin{eqnarray}
     \alpha A_2(x,p) = A_1(x,p) * A_2(x,p) \label{evstar}
\end{eqnarray} 
and (\ref{Abar2}) can be written as,
\begin{eqnarray}
     \alpha\bar{B}_2(x,x) 
     &=& \frac{1}{ (2\pi\hbar)^{n}}\int d^n l \, \bar{B}_1(x,l+x) *_{\theta} \bar{B}_2(l+x,x) \label{Bbar1}
\end{eqnarray} 
Using (\ref{A1bar}), Eqn.(\ref{Abar1}) can be written as,
\begin{align}
     \alpha \bar{B}_2(x-\frac{\hbar}{2}w,x+\frac{\hbar}{2}w) 
     &= \frac{1}{(2\pi\hbar)^{n}}\int d^n l \, \bar{B}_1(x-\frac{\hbar}{2}w,l+x) *_{\theta} \bar{B}_2(l+x,x+\frac{\hbar}{2}w). \label{Bbar2}
\end{align} 
which is essentially the Fourier transform of momentum dependency of (\ref{evstar}). Eqn.(\ref{Bbar1}) and (\ref{Bbar2}) are homogenous integral equations in NC space, with kernels $\bar{B}_1(x,l+x)$ and $\bar{B}_1(x-\frac{\hbar}{2}w,l+x) $ respectively. The kernels are Hermitian as $\bar{B}_1^*(x,l)=\bar{B}_1(l,x)$. 
%
%%%%%%%%%%%%%%%%%%%%%%%%%%%%%%%%%%%%%%%%%%%%%%%%%%%%%%%%%%%%%%%%%%%%%%%
\subsection{Consequences to the Explicit Form of Probability Density}
%%%%%%%%%%%%%%%%%%%%%%%%%%%%%%%%%%%%%%%%%%%%%%%%%%%%%%%%%%%%%%%%%%%%%%%
%
Let $ A_1(x,p) = A_2(x,p) = f(x,p) $ and $ \bar{B}_1(x,l) = \bar{B}_2(x,l) = G(x,l)$, and let $ \alpha = \kappa$ in the preceding subsection. Then Eqn.(\ref{Abar}), (\ref{Bbar2}) and (\ref{Bbar1}) respectively take the forms:
\begin{align}
     G(x-\frac{\hbar}{2}w,\,x+\frac{\hbar}{2} w) &= \int d^n p \, e^{-iw_i p^i} f(x,p) \label{G1}\\
     G(x-\frac{\hbar}{2}w,\,x+\frac{\hbar}{2} w)
     &= \frac{1}{\kappa(2\pi\hbar)^{n}}\int d^n l \, G(x-\frac{\hbar}{2}w,\,l+x) *_{\theta} G(l+x,\,x+\frac{\hbar}{2}w) \label{G2}\\
     G(x,\,x) 
     &= \frac{1}{\kappa (2\pi\hbar)^{n}}\int d^n l \, G(x,\,l+x) *_{\theta} G(l+x,\,x) \label{G3}
\end{align} 
The Eqn.(\ref{G2}) is the integral form of the quantum condition (\ref{*qcond}). 
Integration over $x$ of (\ref{G3}) gives
\begin{eqnarray}
      \int d^n x \, d^n l \, |G(x,l)|^2 = \kappa (2\pi\hbar)^{n}, \label{kernelsi}
\end{eqnarray} 
Therefore, the kernel is also square integrable and hence the integral equations (\ref{G2}) and  (\ref{G3}) are Fredholm integral equations of second kind. The kernel $ G(x, l+x) $ gives rise to at least one eigenvalue equation (\ref{G3}) with eigenvalue $ (1/{\kappa (2\pi\hbar)^{n}}) $ with the corresponding eigenfunction $G(x,x)$. Since the kernel is also Hermitian we can make use of the Hilbert-Schmidt theory which is concerned with the properties of the totality of eigenfunctions and eigenvalues and their connection with kernel and the iterated kernels \cite{Courant}. If $\phi_i(x)$ is an orthonormal sequence (finite or infinite) of eigenfunctions of the kernel, then the kernel  $ G(x, l+x) $ can be developed into the series
\begin{eqnarray}
     G(x,l+x) = \sum_{i=1}^q \frac{\phi_i(x) *_{\theta} \phi^{*}_i(l+x)}{\mu_i} ,\label{ker1} 
\end{eqnarray} 
where $q$ is finite or infinite depending on whether the kernel is degenerate or not. 
Then it follows in particular that
\begin{eqnarray}
     \int d^n x \, d^n l \, |G(x,l)|^2 = \sum_{i=1}^{q} \frac{1}{\mu_i^2},\label{parseval}
\end{eqnarray} 
where $ \mu_i$ such that $ |\mu_1|\leq |\mu_2| \leq \ldots $ are eigenvalues of $ G(x, l+x) $. Comparing (\ref{kernelsi}) and (\ref{parseval}), we get the relation
\begin{eqnarray}
     \frac{1}{\mu_r} = \sum_{i=1}^{r-1} \frac{1}{\mu_i^2} + \frac{m}{\mu_r^2} + \sum_{i=r+m}^{q} \frac{1}{\mu_i^2} ,\label{HSidentity}
\end{eqnarray} 
where $ \mu_r = 1/(\kappa (2\pi\hbar)^{n}) $ and $m$ is the multiplicity in the eigenvalue $\mu_r$. Substituting (\ref{ker1}) in (\ref{G3}), we get  
\begin{align}
         G(x,x) &=  \frac{1}{(\kappa (2\pi\hbar)^{n})} G^{(2)}(x,x)\nonumber \\ 
        &= \mu_r \sum_{i=1}^{q} \frac{\phi_i(x) *_{\theta} \phi^{*}_i(x)}{\mu_i^2} \nonumber
\end{align} 
Moreover, from the relation $ f = \kappa^{1-2N} (f*)^{2N} $, it is easy to show that
\begin{align}
   G(x,x) &= \frac{1}{(\kappa (2\pi\hbar)^{n})^{2N-1}} G^{(2N)}(x,x),  \label{G-GN1} \\
        &= \mu_r^{2N-1} \sum_{i=1}^{q} \frac{\phi_i(x) *_{\theta} \phi^{*}_i(x)}{\mu_i^{2N}} \label{G-GN2}
\end{align} 
where $ G^{(2N)}$ is the $2N$-th iterated kernel
\begin{align}
     G^{(2N)}(x,x) &= \int d^n l_1 \ldots d^n l_{2N-1} G(x, l_1+x) *_{\theta} G(l_1+x, l_2)*_{\theta}  \nonumber \\
                   & \phantom{\int d^n l_1 \ldots d^n l_{2N-1}} \ldots *_{\theta} G(l_{2N-1}+x, x)  \nonumber \\
     &= \sum_{i=1}^{q} \frac{\phi_i(x) *_{\theta} \phi^{*}_i(x)}{\mu_i^{2N}}, \nonumber
\end{align} 
Integration over $x$ on both sides of (\ref{G-GN2}) yields the equality
\begin{eqnarray}
     \frac{1}{\mu_r}\left[\sum_{i=1}^{r-1} \left(\frac{\mu_r}{\mu_i}\right)^{2N} + m + \sum_{i=r+m}^{q} \left(\frac{\mu_r}{\mu_i}\right)^{2N}\right] =1 \label{HSidendity2}
\end{eqnarray} 
As the $|\mu_i|$'s are arranged in ascending order, the third term in (\ref{HSidendity2}) goes to zero as $ N \to \infty$. Since the first term grows indefinitely as $N \to \infty $, $\mu_i$ with $ i\leq r-1 $ are not allowed. Therefore, we get $m/\mu_r$ = 1 (or) $\kappa = (m(2\pi\hbar))^{-1}$. Then, from the relation (\ref{HSidentity}) we deduce that $\mu_i$ with $ i \geq r+m $ are not allowed, and therefore there is only one distinct eigenvalue. The multiplicity $m$ is finite as an eigenvalue can have only finite multiplicity \cite{Courant}. 
Hence the kernel is degenerate. If we identify $m$ with the number of mixed, equally probable and noninterfering states, then for a system in a pure state the kernel is of the form
\begin{eqnarray}
     G(x,l+x) = \phi(x) *_{\theta} \phi^{*}(l+x), \nonumber
\end{eqnarray} 
and because of (\ref{G2}), $G(x-\frac{\hbar}{2}w,x+\frac{\hbar}{2}w)$ becomes
\begin{eqnarray}
     G(x-\frac{\hbar}{2}w,x+\frac{\hbar}{2}w) = \phi(x-\frac{\hbar}{2}w) *_{\theta} \phi^{*}(x+\frac{\hbar}{2}w) ,\label{ker}
\end{eqnarray} 
and the quantum condition (\ref{*qcond})  becomes
\begin{eqnarray}
     f*f = \frac{1}{(2\pi\hbar)^{n}} f. \label{qcondition}
\end{eqnarray} 
The proportionality constant in (\ref{qcondition}) is the same as the one in (\ref{Bakerqcond}). It is the 'expectation value' of $f$ and it characterizes the inverse of 'volume of uncertainty' in phase space. In the limit $ \hbar \to 0 $ in (\ref{Bakerqcond}) the inverse of volume of uncertainty goes to $ 1/0^+$. (Later we will see that the limit $ \hbar \to 0 $ in (\ref{qcondition}) is not possible if $\theta$ is independent of $\hbar$.) It is remarkable that in the classical case the infinities in the relation  $ \delta^{(n)}(x-x_0)\delta^{(n)}(p-p_0)\delta^{(n)}(x-x_0)\delta^{(n)}(p-p_0)= \delta^{(n)}(0)\delta^{(n)}(0) \delta^{(n)}(x-x_0)\delta^{(n)}(p-p_0)$ have a physical meaning. 

Substituting (\ref{ker}) in (\ref{G1}) and taking the inverse Fourier transform of (\ref{G1}) yields 
\begin{eqnarray}
     f(x,p) = \frac{1}{(2\pi)^n} \int d^n w \, e^{i p_i w^i} \phi(x-\frac{\hbar}{2}w) *_{\theta} \phi^{*}(x+\frac{\hbar}{2}w). \label{ncwdf}
\end{eqnarray} 
From (\ref{ncwdf}), it is easy to deduce that
\begin{eqnarray}
     \int d^n x \, d^n p \, f(x,p) = \int d^n x \, \phi^*(x) \phi(x) = 1 \label{normphi}
\end{eqnarray} 
Thus the normalization of $f$ forces that $\phi$ be square integrable and hence belong to a Hilbert space. If we identify $\phi$ with quantum mechanical wave function then (\ref{ncwdf}) is the Wigner distribution function (WDF) in NCQM which is the same as the one in \cite{Jing} and thus corroborates their finding that in NC phasespace the expression for WDF in terms of wave function gets modified. In the limit $ \theta \to 0$, the relation (\ref{ncwdf}) gives the correct expression for WDF of conventional QM:
\begin{eqnarray}
     f_{{\mathrm{c}}}(x,p) = \frac{1}{(2\pi)^n} \int d^n w \, e^{i p_i w^i} \phi(x-\frac{\hbar}{2}w) \phi^{*}(x+\frac{\hbar}{2}w) \label{cwdf}
\end{eqnarray} 
which is in general a nonlocal composition of wave functions. The properties of WDF in the commutative case have been well studied in \cite{Moyal,Baker,Fairlie,Wigner,Hillery,Kruger,Narcowich} Even in the commutative case the WDF can also take negative values [see especially Eqn.(\ref{trace2})]. Hence it is termed as quasi-probability distribution function. 

With little manipulations, Eqn.(\ref{cwdf}) and (\ref{ncwdf}) may also be written as
\begin{eqnarray}
\begin{aligned}
     f_c(x,p) &= \frac{1}{(2\pi)^n} \int d^n w \, \phi(x) *_{\hbar} e^{ip_i w^i}*_{\hbar}  \phi^{*}(x),  \\
     &= \phi(x) *_{\hbar} \delta^{(n)}(p)*_{\hbar}  \phi^{*}(x) \label{cwdf1}
\end{aligned}
\end{eqnarray} 
\begin{eqnarray}
\begin{aligned}
     f(x,p) &= \frac{1}{(2\pi)^n} \int d^n w \, \phi(x) * e^{ip_i w^i}*  \phi^{*}(x), \label{ncwdf1} 
     \\&= \phi(x) * \delta^{(n)}(p) * \phi^{*}(x) 
\end{aligned}
\end{eqnarray} 
Note that WDFs in the the forms (\ref{cwdf1}) and (\ref{ncwdf1}) have not appeared before in the literature. However, we will show later that these forms hold the key to the elegant  establishment of the correspondence between phase space and Hilbert space formulations of QM (NCQM). 
%
%%%%%%%%%%%%%%%%%%%%%%%%%%%%%%%%%%%%%%%%%%%%%%%%%%%%%%%%%%%%%%%%
\section{Variational Methods}
%%%%%%%%%%%%%%%%%%%%%%%%%%%%%%%%%%%%%%%%%%%%%%%%%%%%%%%%%%%%%%%%
\subsection{Quantum-Condition Preserving Variations}
%%%%%%%%%%%%%%%%%%%%%%%%%%%%%%%%%%%%%%%%%%%%%%%%%%%%%%%%%%%%%
%
In this subsection, we look for small possible variations of $f$ and $\phi$ that preserve the quantum condition (\ref{qcondition}) and the normalization (\ref{normal}). In \cite{Baker} Baker pointed out that if $ \delta f = f *_{\hbar} \delta g - \delta g *_{\hbar} f $, for some arbitrary  $\delta g$, then the quantum condition $ f *_{\hbar} f = \frac{1}{(2\pi\hbar)^n} f $ and the normalization of $f$ are preserved. Similar variations preserve (\ref{qcondition}) and (\ref{normal}) in our case. Since $f$ and $\delta f $ are real, $\delta g $ is purely imaginary. Let $\delta g(x,p) = -i h(x,p) $, where $h(x,p)$ is real. Then in our case 
\begin{align}
     \delta f = i [h \scomma f], \label{deltaf}
\end{align}
which preserves the quantum condition (\ref{qcondition}) and the normalization of $f$. Baker, however, did not extend his analysis to the possible variations of $\phi$, which we do because it would establish the link between the phase space and Hilbert space formulations of QM. Since $\phi$ is a function of $x$, we expect $\delta \phi$ to be a function of only $x$; however, because of the explicit form (\ref{ncwdf1}) of $f$ in terms of $\phi$ and because of the following theorem, $\delta\phi$ can also be a function of $p$ in an indirect way. \\

\begin{theorem}
     For two proper functions $ A(x,p)$ and $ B(x)$, the following equality holds. 
\begin{eqnarray}
\begin{aligned}
  \int d^n y \left\{ A(x,p) * B(x) * e^{iy_i p^i} \right\} = \int d^n y \left\{ A(x,-i\hbar  \frac{\partial}{\partial x^i}) *_{\theta} B(x) \right\} * e^{iy_i p^i}, \label{ABet}
 \end{aligned}
\end{eqnarray}
where the derivative $\frac{\partial}{\partial x^i}$ is a differential operator with its Weyl-operator ordering subtleties, and it acts on $x$ dependencies of both $A$ and $B$. The curly bracket in the RHS signifies that the differential operator does not act on $x$-dependent functions, if any, outside the bracket. \\
\end{theorem}
\begin{proof} The proof is based on the method of proof of the Theorem 2 of \cite{Dias1}, where it was shown that the eigenvalue equation in phase space for a physical observable in the case $\theta=0$ leads to the corresponding eigenvalue equation satisfied by an associated wave function in Hilbert space. 

Using the Fourier transforms of $A(x,p)$ and $B(x)$, we write the LHS as
\begin{eqnarray}
     &&\int \frac{d^n k}{(2\pi)^n} \; \frac{d^n l}{(2\pi)^n} \; \frac{d^n k'}{(2\pi)^n} \; d^n y \tilde{A}(k,l) \tilde{B}(k') \left\{  e^{il_i p^i + i k_i x^i} * e^{i k'_i x^i} * e^{ i y_i p^i} \right\}  = \nonumber \\
     && \phantom{\int \frac{d^n k}{(2\pi)^n}} = \int \frac{d^n k}{(2\pi)^n} \; \frac{d^n l}{(2\pi)^n} \; \frac{d^n k'}{(2\pi)^n} \; \tilde{A}(k,l) \tilde{B}(k') E(k,l;k')\label{A2inter0}
\end{eqnarray} 
and consider the  expression
\begin{align}
     E(k,l;k') &= \int d^n y \left\{  e^{il_i p^i + i k_i x^i} * e^{i k_i' x^i} * e^{ i y_i p^i} \right\} \nonumber \\
     &= \int d^n y \left\{ e^{i l_i p^i + i k_i x^i + \frac{\hbar}{2} l_i \overset{\rightarrow}{\partial_{x^i}} - \frac{\hbar}{2} k_i \overset{\rightarrow}{\partial_{p^i}}} *_{\theta} e^{i k_i' x^i -\frac{\hbar}{2} k_i' \overset{\rightarrow}{\partial_{p^i}}}\right\} e^{iy_i p^i} \nonumber \\
     &= \int d^n y \left\{ e^{i k_i x^i + \frac{\hbar}{2} l_i \overset{\rightarrow}{\partial_{x^i}}} *_{\theta}e^{i k_i' x^i} \right\} \left\{ e^{i y_i p^i } e^{l_i \overset{\leftarrow}{\partial_{y^i}}}\right\} \left\{ e^{-\frac{i\hbar}{2}k_i y^i} e^{-\frac{i\hbar}{2}k'_i y^i} \right\}\nonumber \\
     &= \int d^n y \left\{ e^{i k_i x^i + \frac{\hbar}{2} l_i \overset{\rightarrow}{\partial_{x^i}}} *_{\theta}e^{i k_i' x^i} \right\}  e^{i y_i p^i } \left\{ e^{-l_i \overset{\rightarrow}{\partial_{y^i}}} e^{-\frac{i\hbar}{2}k_i y^i} e^{-\frac{i\hbar}{2}k'_i y^i} \right\} \label{A2inter1}
\end{align} 
where in the last step we have done integration by parts in $y$. The Baker-Campbell-Hausdorff relation gives
\begin{eqnarray}
     e^{-l_i \overset{\rightarrow}{\partial_{y^i}}} e^{-\frac{i\hbar}{2}k_i y^i} = e^{-\frac{i\hbar}{2}k_i y^i}e^{-l_i \overset{\rightarrow}{\partial_{y^i}}}e^{-\frac{i\hbar}{2}l_i k_i} \label{A2inter2}
\end{eqnarray} 
Substituting (\ref{A2inter2}) in (\ref{A2inter1}), we get
\begin{align}
      E &=\int d^n y \left\{ e^{i k_i x^i + \frac{\hbar}{2} l_i \overset{\rightarrow}{\partial_{x^i}}} *_{\theta}e^{i k_i' x^i} \right\}  e^{i y_i p^i } \left\{  e^{-\frac{i\hbar}{2}k_i y^i}e^{-\frac{i\hbar}{2}l_i k_i}e^{-l_i \overset{\rightarrow}{\partial_{y^i}}} e^{-\frac{i\hbar}{2}k'_i y^i} \right\} \nonumber \\
      &= \int d^n y \; e^{i y_i p^i } \left\{ e^{ik_i( x^i -\frac{\hbar}{2} y^i )} e^{ l_i (\frac{\hbar}{2} \overset{\rightarrow}{\partial_{x^i}}- \overset{\rightarrow}{\partial_{y^i}})} e^{\frac{i\hbar}{2} l_i k^i} *_{\theta} e^{ik_i' (x^i-\frac{i\hbar}{2}y^i)} \right\}   \nonumber \\
      &= \int d^n y \; e^{i y_i p^i } \left\{ e^{ik_i( x^i -\frac{\hbar}{2} y^i ) + l_i (\frac{\hbar}{2} {\partial_{x^i}}- {\partial_{y^i}})} *_{\theta} e^{ik_i' (x^i-\frac{\hbar}{2}y^i)} \right\} \label{A2inter3},
\end{align} 
where in the last step we have made use of the Baker-Campbell-Hausdorff relation $ e^X e^Y e^{-\frac{1}{2} [X,Y]} = e^{ X + Y}$ with $ X = ik_i( x^i -\frac{\hbar}{2} y^i ) $ and $ Y= l_i (\frac{\hbar}{2} \overset{\rightarrow}{\partial_{x^i}}- \overset{\rightarrow}{\partial_{y^i}}) $. We note that for some functions $f(x)$ and $g(x)$, 
\begin{eqnarray}
     \left\{f(\partial_{y^i}) g(x^i-\frac{\hbar}{2}y^i)\right\} = \left\{f(-\frac{\hbar}{2} \partial_{x^i}) g(x^i-\frac{\hbar}{2}y^i)\right\} \label{A2inter4}
\end{eqnarray} 
Using (\ref{A2inter4}) in (\ref{A2inter3}), we get
\begin{align}
     E &=\int d^n y \; e^{i y_i p^i } \left\{ e^{ik_i( x^i -\frac{\hbar}{2} y^i ) + il_i (-i\hbar {\partial_{x^i}} )} *_{\theta} e^{ik_i' (x^i-\frac{\hbar}{2}y^i)} \right\} \nonumber\\
     &=\int d^n y \; \left\{ e^{ik_i x^i  + il_i (-i\hbar {\partial_{x^i}} )} *_{\theta} e^{ik_i' x^i} \right\} *_{\hbar} e^{i y_i p^i } \nonumber \\
     &=\int d^n y \; \left\{ e^{ik_i x^i  + il_i (-i\hbar {\partial_{x^i}} )} *_{\theta} e^{ik_i' x^i} \right\} * e^{i y_i p^i } \label{A2inter5}
\end{align} 
Substituting (\ref{A2inter5}) in (\ref{A2inter0}), we get
\begin{eqnarray}
\begin{aligned}
     \int d^n y \{ A(x,p) * B(x) * e^{iy_i p^i} \} =& \int \frac{d^n k}{(2\pi)^n} \; \frac{d^n l}{(2\pi)^n} \; \frac{d^n k'}{(2\pi)^n} \; d^n y \tilde{A}(k,l) \tilde{B}(k') \times\\ 
     & \phantom{\frac{d^n k}{(2\pi)}} \times \left\{ e^{ i k_i x^i + i l_i \left( -i \hbar \frac{\vec{\partial}}{\partial x^i}\right)} *_{\theta} e^{ i k_i' x^i } \right\} * e^{i y_i p^i}\\
     =& \int d^n y \left\{ A(x,-i\hbar \frac{\partial}{\partial x^i}) *_{\theta} B(x) \right\} * e^{iy_i p^i}.  \; \blacksquare  \label{ABe1}
\end{aligned}
\end{eqnarray} 
\end{proof}
\begin{corollary}
For real $A(x,p)$,
\begin{eqnarray}
\begin{aligned}
  \int d^n y  \left\{ e^{iy_i p^i} * B^*(x) * A(x,p) \right\} = \int d^n y \; e^{iy_i p^i} * \left\{ B^*(x) *_{\theta} A(x,i\hbar \frac{\overset{\leftarrow}{\partial}}{\partial x^i}) \right\} \label{eBAt}
 \end{aligned}
\end{eqnarray}
\end{corollary}
\begin{proof} This follows immediately by taking the complex conjugate of (\ref{ABe1}).  $\blacksquare$ \\
\end{proof}

\begin{corollary} For two phase space functions $ A_1(x,p)$ and $ A_2(x,p)$, the following equality holds:

\begin{eqnarray}
\begin{aligned}
  & \int d^n y \left\{ A_1(x,p)*A_2(x,p) * B(x) * e^{iy_i p^i} \right\} = \\&= \int d^n y \left\{ A_1(x,-i\hbar  \frac{\partial}{\partial x^i}) *_{\theta}A_2(x,-i\hbar  \frac{\partial}{\partial x^i}) *_{\theta} B(x) \right\} * e^{iy_i p^i}, \label{T2C2}
 \end{aligned}
\end{eqnarray}
\end{corollary}
\begin{proof}
This follows immediately by taking $ A(x,p) = A_1(x,p) *A_2(x,p) $ in (\ref{ABe1}).  $\blacksquare$ \\
\end{proof}

Note that Corollary 2 establishes the correspondence between the star product and the operator product. \\ 
\begin{theorem}
\label{variphi}
If $ \phi \to \phi + i \left( h(x^i, -i\hbar \frac{\partial}{\partial x^i}) *_{\theta} \phi \right) $, then $\delta f$ satisfies (\ref{deltaf}). \\
\end{theorem}
\begin{proof}
From (\ref{ncwdf1}) it follows that
\begin{align}
     \delta f &= \frac{1}{(2\pi)^n} \int d^n w \, \left[ \delta \phi(x) * e^{ip_i w^i}*  \phi^{*}(x) + \phi(x) * e^{ip_i w^i}*  \delta \phi^{*}(x) \right] \nonumber\\
     &= \frac{1}{(2\pi)^n} \int d^n w \, \left[   \left\{ih(x,\, -i\hbar \frac{\overset{\rightarrow}{\partial}}{\partial x^i}) *_{\theta} \phi(x)\right\} * e^{ip_i w^i}*  \phi^{*}(x) \right. - \nonumber \\
     & \left. \phantom{\frac{1}{(2\pi)^n} \int d^n w \frac{1}{(2\pi)^n} } - \phi(x) * e^{ip_i w^i}*   \left\{\phi^{*}(x) *_{\theta} ih(x,\, i\hbar \frac{\overset{\leftarrow}{\partial}}{\partial x^i})\right\} \right] \nonumber \\
     &= \frac{1}{(2\pi)^n} \int d^n w \, \left\{ i  (h(x,\,p) * \phi(x)) * e^{ip_i w^i}*  \phi^{*}(x) \right. - \nonumber \\ 
     & \phantom{\frac{1}{(2\pi)^n} \int d^n w}\left.  - \phi(x) * e^{ip_i w^i}*  (\phi^{*}(x) * ih(x,\,p)) \right\} \label{deltaf1}
\end{align} 
In the last step, we have made use of (\ref{ABet}) and (\ref{eBAt}). By the associativity property of star product, the proof follows simply from (\ref{deltaf1}). $\blacksquare$ \\
\end{proof}
Note, however, that $h(x,p) * \phi(x) \neq \left( h(x,-i\hbar \frac{\partial}{\partial x^i}) *_{\theta} \phi(x) \right) $. The equality holds only under the integration of the form (\ref{ABe1}).   
%
%%%%%%%%%%%%%%%%%%%%%%%%%%%%%%%%%%%%%%%%%%%%%%%%%%%%%%%%%%%%
\subsection{Extremizing the Expectation Value}
%%%%%%%%%%%%%%%%%%%%%%%%%%%%%%%%%%%%%%%%%%%%%%%%%%%%%%%%%%%%
%
The expectation value of a physical observable $ A(x,p) $ is
\begin{eqnarray}
     \left< A \right> &=& \frac{\int\; d^n x \; d^n p\; A(x,p)* f(x,p) }{\int \; d^n x \; d^n p\; f(x,p)} \label{expect1}
\end{eqnarray} 
We insert the Fourier synthesis of unity (\ref{unity}) of momentum space
\begin{eqnarray}
    1 = \int d^n w \; \delta^n(w) e^{-iw_ip^i} \label{unity}
\end{eqnarray} 
in the integrand of the numerator to get
\begin{eqnarray}
     \left< A \right> &=& \frac{\int\; d^n x \; d^n p\;\int d^n w \; \delta^n(w) e^{-iw_ip^i}  \left\{A(x,p)* f(x,p)\right\} }{\int \; d^n x \; d^n p\; f(x,p)} \nonumber
\end{eqnarray} 
Identifying $A$ with $A_1$ and $f$ with $ A_2 $ of (\ref{A1*A2}) and using the RHS of (\ref{Bbar2}) and the expression (\ref{ncwdf}), the above expression is written as
\begin{align}
     \left< A \right> &= \frac{\int\; d^n x d^n w \; \delta^n(w) \int d^n l \, \bar{B}_1(x-\frac{\hbar}{2}w,l+x) *_{\theta} G(l+x,x+\frac{\hbar}{2}w)}{(2\pi\hbar)^n \int d^n x \, d^n l d^n w \delta^n(w) \,G(x-\frac{\hbar}{2}w,\,x+\frac{\hbar}{2}w) } \nonumber \\
     &= \frac{\int\; d^n x d^n w \; \delta^n(w) \int d^n l \, \left\{ \bar{B}_1(x-\frac{\hbar}{2}w,l+x) *_{\theta} \phi(l+x) *_{\theta} \phi^*(x+\frac{\hbar}{2}w) \right\} }{(2\pi\hbar)^n \int d^n x \; d^n l \;d^n w \;\delta^n(w) \;\phi(x-\frac{\hbar}{2}w)\phi^*(x+\frac{\hbar}{2}w) } \nonumber \\
     &= I[\phi] \label{expect3}
\end{align}
where $\bar{B}_1$ defined in (\ref{A1bar}) is related to the Fourier transform of momentum dependency of $ A(x,p)$. For notational simplicity of the analysis, we define the operator ${\mathcal B}$ and an inner product $\left<\; , \;\right>$ as follows:
\newcommand{\mb}{{\mathcal B}}
\begin{align}
     \mb &= \frac{1}{(2\pi\hbar)^n}\int d^n l \,  \bar{B}_1(x-\frac{\hbar}{2}w,l+x) *_{\theta}   \nonumber\\
     \left< \phi , \phi \right> &= \int d^n x \; d^n l \;d^n w \;\delta^n(w)  \;\phi(x-\frac{\hbar}{2}w)*_{\theta} \phi^*(x+\frac{\hbar}{2}w) \nonumber
\end{align} 
Then the expectation value (\ref{expect3}) is written as
\begin{align}
     I[\phi] = \frac{\left< \mb\phi , \phi \right>}{\left< \phi , \phi \right>} = \frac{\left< \phi , \mb\phi \right>}{\left< \phi , \phi \right>} \nonumber
\end{align} 
For arbitrarily small variations $\delta\phi$, 
\begin{align*}
     I[\phi +\delta\phi] &= \frac{\left< \mb\phi+\delta\phi , \phi+\delta\phi \right>}{\left< \phi+\delta\phi , \phi+\delta\phi \right>} \\
     &= \frac{\left< \mb\phi , \phi \right> + \left< \mb\delta\phi , \phi \right>+ \left< \mb\phi , \delta \phi \right>+ \left< \mb\delta\phi , \delta\phi \right>}{\left< \phi , \phi \right>\left[ 1+ \frac{\left< \delta\phi , \phi \right>}{\left< \phi , \phi \right>}+ \frac{\left< \phi , \delta\phi \right>}{\left< \phi , \phi \right>}+ \frac{\left< \delta\phi , \delta\phi \right>}{\left< \phi , \phi \right>}\right]} \\
     &= I[\phi] + \frac{\left< \mb\delta\phi , \phi \right>+ \left< \mb\phi , \delta \phi \right>}{\left< \phi , \phi \right>} - \frac{\left< \mb\phi , \phi \right>}{\left< \phi , \phi \right>^2 } \left\{\left< \delta\phi , \phi \right> + \left< \phi , \delta\phi \right>\right\} + {\mathcal O}[(\delta\phi)^2]
\end{align*} 
Demanding that $I[\phi]$ be stationary under such small changes and calling $ \lambda$ the stationary value of $I[\phi]$, we get
\begin{eqnarray}
     \left< \delta\phi , (\mb\phi -\lambda\phi ) \right>+ \left< (\mb\phi - \lambda\phi), \delta \phi \right> = 0 . \label{excond}
\end{eqnarray} 
For quantum-condition preserving $ \delta\phi(x) = i \left\{ h(x^i, -i\hbar \frac{\partial}{\partial x^i})  * \phi(x)\right\} $, the above equality gives two equations:
\begin{eqnarray}
     \delta^n(w) \left\{\frac{1}{(2\pi\hbar)^n} \int d^n l \, \bar{B}_1(x-\frac{\hbar}{2}w,l+x) *_{\theta} \phi(l+x) - \lambda \phi(x-\frac{\hbar}{2}w) \right\} = 0, \label{evexcond}
\end{eqnarray} 
and its ``complex conjugate''
\begin{eqnarray}
     \delta^n(w) \left\{\frac{1}{(2\pi\hbar)^n} \int d^n l \,   \phi^*(l+x)*_{\theta} \bar{B}_1(l+x,x+\frac{\hbar}{2}w) - \lambda \phi^*(x+\frac{\hbar}{2}w) \right\} = 0. \label{evexcond2}
\end{eqnarray} 

From (\ref{evexcond}) it is clear that at $ w= 0 $ the following equation holds:
\begin{eqnarray}
     \frac{1}{(2\pi\hbar)^n} \int d^n l \, \bar{B}_1(x,l+x) *_{\theta} \phi(l+x) = \lambda \phi(x) \label{HSev1}
\end{eqnarray} 
Eq.(\ref{evexcond}) is mute as to what happens to the function in the curly bracket in (\ref{evexcond}) at points $ w \neq 0 $. However, Eq.(\ref{HSev1}) is valid for all points in the position space. The change of coordinate $x \to x-\frac{\hbar}{2}w$ in (\ref{HSev1}) gives 
\begin{eqnarray}
     \frac{1}{(2\pi\hbar)^n} \int d^n l \, \bar{B}_1(x-\frac{\hbar}{2}w,l+x-\frac{\hbar}{2}w) *_{\theta} \phi(l+x-\frac{\hbar}{2}w) = \lambda \phi(x-\frac{\hbar}{2}w)  \label{PoS}
\end{eqnarray} 
Right $*_{\theta}$-multiplication on both sides of (\ref{PoS}) by $\phi^*(x+\frac{\hbar}{2}w) $ and the change of integration variable $l-\frac{\hbar}{2}w \to l $ gives,
\begin{align}
     \frac{1}{(2\pi\hbar)^n} \int d^n l \, \bar{B}_1(x-\frac{\hbar}{2}w,l+x) *_{\theta} G(l+x,\, x+\frac{\hbar}{2}w) = \lambda G(x-\frac{\hbar}{2}w, \, x+\frac{\hbar}{2}w) . \label{PSev1}
\end{align} 
From (\ref{evexcond2}), it follows that
\begin{eqnarray}
     \frac{1}{(2\pi\hbar)^n} \int d^n l \,   \phi^*(l+x)*_{\theta} \bar{B}_1(l+x,x) = \lambda \phi^*(x) \label{HSev2}
\end{eqnarray} 
which is merely the complex conjugate of (\ref{HSev1}) and doesn't have any further significance. However, at points $ w \neq 0 $, we get 
\begin{eqnarray}
     \frac{1}{(2\pi\hbar)^n} \int d^n l \,   \phi^*(l+x)*_{\theta} \bar{B}_1(l+x,x+\frac{\hbar}{2}w) = \lambda \phi^*(x+\frac{\hbar}{2}w) \nonumber
\end{eqnarray} 
which is a coordinate-shifted form of (\ref{HSev2}) (followed by a change of integration variable). The left $*_{\theta}$-multiplication by $\phi(x-\frac{\hbar}{2}w)$ gives
\begin{align}
     \frac{1}{(2\pi\hbar)^n} \int d^n l \,  G(x-\frac{\hbar}{2}w, \, l+x)*_{\theta} \bar{B}_1(l+x,x+\frac{\hbar}{2}w) = \lambda G(x-\frac{\hbar}{2}w, \, x+\frac{\hbar}{2}w)  \label{PSev2}
\end{align} 
If the kernel $\bar{B}_{1}$  of (\ref{HSev1}) is square integrable then (\ref{HSev1}) is a noncommutative Fredholm integral equation of second kind. Such an equation possesses at least one eigenvalue and possibly denumerably infinite eigenvalues $\{\lambda_i\}$, and the set of eigenfunctions $\{\psi_i(x)\}$ forms a complete orthonormal set \cite{Courant}. 
Although the Eqn.(\ref{PSev2}) is merely the complex conjugate of (\ref{PSev1}) followed by a sign change in $w$, it has a deeper significance than that of (\ref{HSev2}). We will come to this point in the next Section. 

It's worth remarking that (\ref{PoS}) is the point of separation of two different but equivalent descriptions of quantum mechanics in Hilbert space and in phase space. While the right $*_{\theta}$-multiplication of (\ref{PoS}) by $\phi^*(x+\frac{\hbar}{2}w)$ leads through (\ref{PSev2}) to phase space formulation of stationary state description of NCQM, left  $*_{\theta}$-multiplication by $\phi^*(x-\frac{\hbar}{2}w) $ leads to the same in Hilbert space. 
%
%%%%%%%%%%%%%%%%%%%%%%%%%%%%%%%%%%%%%%%%%%%%%%%%%%%%%%%%%%
\section{Eigenvalue Problems} 
%%%%%%%%%%%%%%%%%%%%%%%%%%%%%%%%%%%%%%%%%%%%%%%%%%%%%%%%%%
\subsection{Eigenvalue Problem in Hilbert Space}
%%%%%%%%%%%%%%%%%%%%%%%%%%%%%%%%%%%%%%%%%%%%%%%%%%%%
%
In this Subsection, we derive an operator-valued eigenvalue equation from the homogeneous integral equation (\ref{HSev1}).  One can also start from (\ref{PSev1}) which is merely the coordinate shifted form of (\ref{HSev1}). In terms of the Fourier transformation $ \bar{A}(x,l) $ of momentum dependency of $ A(x,p)$, Eqn.(\ref{HSev1}) is given by  
\begin{align}
     \lambda \phi(x)&=\frac{1}{(2\pi\hbar)^n} \int d^n l \, \bar{A}\left((x+\frac{1}{2}l),\; \frac{l}{\hbar}\right) *_{\theta} \phi(l+x)  \label{HSev3} \\
     &= \frac{1}{(2\pi\hbar)^n} \int d^n l \, d^n p \, e^{-\frac{i}{\hbar} l_i p^ i} A\left((x+\frac{1}{2}l),\; p\right) *_{\theta} \phi(l+x) \nonumber \\
     &= \frac{1}{(2\pi)^n} \int \frac{d^n k}{(2\pi)^n}\frac{d^n w}{(2\pi)^n} \frac{d^n k'}{(2\pi)^n} d^n y \, d^n p \, \tilde{A}(k,w) \tilde{\phi}(k') \times \nonumber \\
     & \phantom{\frac{1}{(2\pi\hbar)^n}\times \int d^n l } \times \left\{ e^{ik_i(x^i+\frac{\hbar}{2} y^i) + iw_i p^i} *_{\theta} e^{ik_i' ( x^i + \hbar y^i)} \right\} e^{-y_i p^ i}  \nonumber \\
     &= \frac{1}{(2\pi)^n} \int \frac{d^n k}{(2\pi)^n}\frac{d^n w}{(2\pi)^n} \frac{d^n k'}{(2\pi)^n} \tilde{A}(k,w) \tilde{\phi}(k') E(k,w;k') \label{HSevinter1}
\end{align}
where in the third step we have done the change of integration variable $ y= l/\hbar$ and expressed $A$ and $\phi$ in terms of their Fourier transforms. Next, consider the integral expression $E(k,w;k')$:
\begin{align}
     E &= \frac{1}{(2\pi)^n} \int d^n y \, d^n p \, \left\{ e^{ik_i(x^i+\frac{\hbar}{2} y^i) + iw_i p^i} *_{\theta} e^{ik_i' ( x^i + \hbar y^i)} \right\} e^{-y_i p^ i} \nonumber\\
     &= \frac{1}{(2\pi)^n} \int d^n y \, d^n p \, \left\{ e^{ik_i x^i}*_{\theta} e^{i k_i' x^i} \right\} \left\{ e^{\frac{i\hbar}{2} k_i y^i} e^{i\hbar k_i' y^i} e^{-w_i \vec{\partial}_{y^i}} e^{-iy_i p^i} \right\} \nonumber \\
     &= \frac{1}{(2\pi)^n} \int d^n y \, d^n p \, \left\{ e^{ik_i x^i}*_{\theta} e^{i k_i' x^i} \right\} e^{-iy_i p^i} \left\{ e^{w_i \vec{\partial}_{y^i}} e^{\frac{i\hbar}{2} k_i y^i}e^{i\hbar k_i' y^i} \right\} \nonumber  \\
     &= \frac{1}{(2\pi)^n} \int d^n y \, d^n p \, \{ e^{ik_i x^i}*_{\theta} e^{i k_i' x^i} \} e^{-iy_i p^i}  \{ e^{\frac{i\hbar}{2} k_i y^i} e^{\frac{i\hbar}{2} k_i w_i} e^{w_i \vec{\partial}_{y^i}} e^{i\hbar k_i' y^i} \} \nonumber \\
     &= \frac{1}{(2\pi)^n} \int d^n y \, d^n p \, e^{-iy_i p^i} \{ e^{ik_i x^i} e^{\frac{i\hbar}{2} k_i y^i} e^{\frac{i\hbar}{2} k_i w_i} e^{w_i \vec{\partial}_{y^i}} *_{\theta} e^{i k_i' (x^i + \hbar y^i)}\} \label{HSevinter2}
\end{align} 
where in the third step, we have done integration by parts in $y$, and in the fourth step we have used the Baker-Campbell-Hausdorff relation $ e^{w_i \vec{\partial}_{y^i}} e^{\frac{i\hbar}{2} k_i y^i} = e^{\frac{i\hbar}{2} k_i y^i} e^{\frac{i\hbar}{2} k_i w_i} e^{w_i \vec{\partial}_{y^i}} $. Further, we note that $ e^{w_i \vec{\partial}_{y^i}}e^{i k_i' (x^i + \hbar y^i)} = e^{\hbar w_i \vec{\partial}_{x^i}} e^{i k_i' (x^i + \hbar y^i)}  $ and by the Baker-Campbell-Hausdorff relation that 
\begin{align}
 e^{ik_i x^i}e^{\frac{i\hbar}{2} k_i w_i}e^{\hbar w_i \vec{\partial}_{x^i}} = e^{ik_i x^i +  \hbar w_i \partial_{x^i}} .
\end{align}
Then (\ref{HSevinter2}) boils down to 
\begin{align}
     E &= \frac{1}{(2\pi)^n} \int d^n y \, d^n p \, e^{-iy_i p^i} \left\{ e^{ik_i (x^i+\frac{\hbar}{2} y^i) +  \hbar w_i \partial_{x^i} }  *_{\theta} e^{i k_i' (x^i + \hbar y^i)} \right\} \label{HSevinter3}
\end{align} 
Substituting (\ref{HSevinter3}) in (\ref{HSevinter1}), we get
\begin{align}
     \lambda\phi &= \frac{1}{(2\pi)^n} \int \frac{d^n k}{(2\pi)^n}\frac{d^n w}{(2\pi)^n} \frac{d^n k'}{(2\pi)^n} d^n y \, d^n p \, \tilde{A}(k,w) \tilde{\phi}(k') \times \nonumber \\
      & \phantom{\frac{1}{(2\pi)^n} \frac{1}{(2\pi)^n}} \times \left\{ e^{ik_i (x^i+ \frac{\hbar}{2} y^i) + \hbar w_i \partial_{x^i} }  *_{\theta} e^{i k_i' (x^i + \hbar y^i)} \right\} e^{-y_i p^ i} \nonumber\\
     &= \int \frac{d^n k}{(2\pi)^n}\frac{d^n w}{(2\pi)^n} \frac{d^n k'}{(2\pi)^n}  \, \tilde{A}(k,w) \tilde{\phi}(k') \left\{ e^{ik_i x^i+ \hbar w_i \partial_{x^i} }  *_{\theta} e^{i k_i' x^i} \right\}  \nonumber \\
     &= A(x, -i\hbar \frac{\partial}{\partial x^i}) *_{\theta} \phi(x) \nonumber
\end{align} 
which is the differential form of the integral equation (\ref{HSev1}). 
%%%%%%%%%%%%%%%%%%%%%%%%%%%%%%%%%%%%%%%%%%%%%%%%%%%%%%%%%%%%%%%%%%
\subsection{Eigenvalue Problem in Phase Space}   
%%%%%%%%%%%%%%%%%%%%%%%%%%%%%%%%%%%%%%%%%%%%%%%%%%%%%%%%%%%%%%%%%%
%
In this Subsection, we obtain from (\ref{PSev1}) and (\ref{PSev2})  eigenvalue equations in phase space for a physical observable $ A(x,p)$. For the simplicity of the analysis, we consider here the square integrable kernel because the orthonormal and completeness properties of its eigenfunctions are readily known. Let us denote by $ G_i(x-\frac{\hbar}{2}w, \, x+\frac{\hbar}{2}w)$ the eigenfunction of (\ref{PSev1}) with an eigenvalue $ \lambda_i$. Each such $G_i$  can be written in the form 
\begin{eqnarray}
     G_i(x-\frac{\hbar}{2}w, \, x+\frac{\hbar}{2}w) = \psi_i(x-\frac{\hbar}{2}w) *_{\theta} \phi^*(x+\frac{\hbar}{2}w), \label{G_a1}
\end{eqnarray} 
where $ \psi_i(x-\frac{\hbar}{2}w)$ is an eigenfunction of (\ref{PoS}). Moreover, let $G_j(x+\frac{\hbar}{2}w, \, x-\frac{\hbar}{2}w)$ be an eigenfunction of (\ref{PSev2}) and it can be written as 
\begin{eqnarray}
     G_j(x+\frac{\hbar}{2}w, \, x-\frac{\hbar}{2}w) = \phi(x+\frac{\hbar}{2}w) *_{\theta} \psi_j^*(x-\frac{\hbar}{2}w),  \label{G_a2}
\end{eqnarray} 
where $\psi_j^*(x-\frac{\hbar}{2}w)$ is the corresponding eigenfunction of (\ref{HSev2}). Then the set $\{G_{ij}(x,x)\} = \{\psi_i(x) *_{\theta} \psi_j^*(x)\} $ forms the common set of eigenfunctions of (\ref{PSev1}) and (\ref{PSev2}) at $w =0$. However, for the extremal condition (\ref{excond}) to be satisfied, both (\ref{PSev1}) and (\ref{PSev2}) should hold simultaneously. That is, if $\lambda_i$ is the value  that extremizes (\ref{expect3}) through (\ref{excond}), then both (\ref{PSev1}) and (\ref{PSev2}) should have the same eigenvalue simultaneously. This can happen only by the subset $\{G_{ii}(x,x)\}$ of $\{G_{ij}(x,x)\}$. Therefore, not all common eigenfunctions $G_{ij}(x,x)$ can extremize the functional (\ref{expect3}). This is the subtle deviation from the conventional theory of integral equations. The deviation is caused by the extra integral equation (\ref{G2}) that the eigenfunctions should obey. The integral equation (\ref{G2}), which is nothing but the Fourier transform of momentum dependency of the pure state quantum condition, splits $G$ into two independent constituent functions as in (\ref{ker}) which lead to two different conditions (\ref{PSev1}) and (\ref{PSev2}) that should hold simultaneously. 

Next, we consider the inverse Fourier transform of (\ref{PSev1}) and (\ref{PSev2}). Note that (\ref{Bbar2}) is the Fourier transform of momentum dependency of (\ref{evstar}). In other words, (\ref{evstar}) is the inverse Fourier transform of (\ref{Bbar2}). In the same way, the inverse Fourier transforms of (\ref{PSev1}) and (\ref{PSev2}) turn out to be
\begin{eqnarray}
     A(x,p) * f(x,p) - \lambda f(x,p) = 0; \qquad f(x,p) * A(x,p) - \lambda f(x,p) = 0. \label{PSev}
\end{eqnarray} 
Each $G_{ij}$ in the common set $\{G_{ij}\}$ of eigenfunctions of (\ref{PSev1}) and (\ref{PSev2}) gives rise to a corresponding $f$ which may be denoted by $f_{ij}$. Since both of the above equations should hold simultaneously to extremize (\ref{expect3}), only the diagonal subset $\{f_{ii}\}$ corresponds to observable phase space states. However, it can be shown that it is the common set $\{f_{ij}\}$ that forms a complete orthonormal set of eigenfunctions of (\ref{PSev}) if the set $\{\psi_i(x)\}$ is complete. Since $G_{ij}$ is Hermitian $f_{ij}$ is also Hermitian i.e., $f_{ij}(x,p) = f_{ji}^{*}(x,p)$. 

For the special case in which $A$ is the Hamiltonian, Eqn.(\ref{PSev}) was explicitly worked out in \cite{Jing} using NC WDF and the assumption that the Hamiltonian operator satisfies the eigenvalue equation in Hilbert space. The properties of corresponding $f_{ij}$ were also given there. For the sake of completeness, we provide here those properties as they are also applicable to  the case in which the kernel $\bar{B}_{1}$ in (\ref{PSev1}) is square integrable.  To sum up, if the set $\{\psi_i(x)\}$ forms a complete orthonormal set, then $f_{ij}$ have the following properties \cite{Moyal,Fairlie,Jing}:
\begin{align}
     %f_{ij}(x,p) &= f_{ji}^*(x,p)  \nonumber \\
     \int d^n x \, d^n p \, f_{ij}(x,p) &= \delta_{ij} \nonumber \\
     (2\pi\hbar)^n\sum_{i} f_{ii}(x,p) &= 1 \nonumber \\
     f_{ij}(x,p) * f_{kl}(x,p) &= \frac{1}{(2\pi\hbar)^n} \delta_{jk} f_{il}(x,p)  \label{qcondij} \\
     \int d^n x \, d^n p \, f_{ij}(x,p) f^*_{kl}(x,p) &= \frac{1}{(2\pi\hbar)^n} \delta_{ik} \delta_{jl} \label{trace2}\\
     \sum_{ij} f_{ij}(x,p) f^*_{ij}(x',p') &= \frac{1}{(2\pi\hbar)^n} \delta^{(n)}(x-x') \delta^{(n)}(p-p') \label{complete2}
\end{align}
In the case $ \theta =0 $, except (\ref{qcondij}) all other properties were worked out in  \cite{Moyal}, and the property (\ref{qcondij}) was first derived in \cite{Fairlie} from the eigenvalue equation of a Hamiltonian. From the properties (\ref{trace2}) and (\ref{complete2}) it follows that, not only do the $ f_{ij}$ form a basis for the quasi-probability distribution functions, but they also span the entire Hilbert space of functions on phase space. 
%
%%%%%%%%%%%%%%%%%%%%%%%%%%%%%%%%%%%%%%%%%%%%%%%%%%%%%%%%%%%%%%%%%
\section{Dynamical Equations}
%%%%%%%%%%%%%%%%%%%%%%%%%%%%%%%%%%%%%%%%%%%%%%%%%%%%%%%%%%%%%%%%%
\subsection{Generalized Moyal Equation of Motion}
%%%%%%%%%%%%%%%%%%%%%%%%%%%%%%%%%%%%%%%%%%%%%%%%%%%%%%%%
%
In this section, we consider the dynamical aspect of the theory. In the Hilbert space to phase space approach, the dynamical equation is obtained by taking the Wigner-transform of von Neumann equation \cite{Moyal,Eftekharzadeh}. Within the postulational phase space formulation, it can be obtained by considering the quantum-condition preserving time variation of $f$ and relying only on the classical limit without any recourse to QM in Hilbert space \cite{Baker}. We work it out for the case of NCQM. A small variation in the expectation value of momentum $p_i$ [cf. (\ref{expect})] is given by 
\begin{eqnarray}
     \frac{d \left< p \right>}{d t} \delta t = \int  d^n x \, d^n p \, p_i \frac{\partial f(x,p,t)}{\partial t} \delta t \label{expectp}
\end{eqnarray}
Preserving the normalization and quantum condition, the change in $ f$ may also be written as [cf. (\ref{deltaf})]. %
\begin{eqnarray}
     \delta f = \left(\partial f/\partial t\right) \delta t = i[h \scomma f] .\label{Moyal1}
\end{eqnarray} 
Then in the limit $\hbar \to 0$, the RHS of (\ref{expectp}) becomes
\begin{align}
     \int  d^n x d^n p  p_i \frac{\partial f(x,p,t)}{\partial t} \delta t 
     &= i\int  d^n x  d^n p [p_i \scomma h] f \nonumber \\
     &= i\int  d^n x  d^n p  i\hbar \left\{p_i, h\right\} f\nonumber \\
     &= \int  d^n x  d^n p  \left\{p_i, H\right\} f \delta t,\label{deqn2}
\end{align} 
where in the last step we have made use of the correspondence principle; $\{\}$ is the classical Poisson bracket and $H$ is the Hamiltonian. Since the Eqn.(\ref{deqn2}) must hold for all $p_i$, we have  $ h = -(1/\hbar)H \delta t $. Then from (\ref{Moyal1}), we get the phase space version of von Neumann equation or the generalized Moyal equation of motion:
\begin{eqnarray}
      \frac{\partial f}{\partial t} = \frac{1}{i\hbar}[H \scomma f] .\label{Moyal2}
\end{eqnarray} 
It's worth remarking that the limit $\theta \neq 0$ and $ \hbar \to 0$ of the RHS of (\ref{Moyal2}) does not consistently exist.  That is, nonrelativistic classical mechanics with $*_{\theta}$ deformation in the spatial directions is not a limiting case of a quantum theory.  This result was obtained in \cite{Eftekharzadeh}. However, if $\theta_{ij}$ is of the form $\hbar\bar{\theta}_{ij}$, then the limit exists and we have a classical theory with nontrivial Poisson bracket between spatial coordinates. In other words, the Poisson bracket in this case becomes
\begin{align}
     \{A,B\} = \left( \frac{\partial A}{\partial x^i} \frac{\partial B}{\partial p_i} -\frac{\partial A}{\partial p_i} \frac{\partial B}{\partial x^i}\right) + \bar{\theta}^{ij} \frac{\partial A}{\partial x^i} \frac{\partial B}{\partial x^j} \nonumber
\end{align} 
Such a modified Poisson bracket has already been employed in the classical context \cite{Romero}. 
%
%%%%%%%%%%%%%%%%%%%%%%%%%%%%%%%%%%%%%%%%%%%%%%%%%%%%%%%%%%%%%%%%%%
\subsection{Time dependent Schr\"{o}dinger Equation}
%%%%%%%%%%%%%%%%%%%%%%%%%%%%%%%%%%%%%%%%%%%%%%%%%%%%%%%%%%%%%%%%%%
%
The crucial point in the preceding analysis is that, to be in touch with classical mechanics,   $h$  in (\ref{Moyal1}) should be equal to $ -(1/\hbar)H \delta t $. Such a variation is produced by the following form of $\delta\phi$ (see Theorem \ref{variphi}): 
\begin{eqnarray}
     \delta \phi = \delta t \frac{\partial\phi}{\partial t}&=& \delta t\frac{1}{i\hbar} H(x^i, -i\hbar\frac{\partial}{\partial x^i}) *_{\theta} \phi ,\nonumber
\end{eqnarray} 
from which we get the  Schr\"{o}dinger equation:
\begin{align}
     i\hbar \frac{\partial\phi}{\partial t}=  H (x^i, -i\hbar\frac{\partial}{\partial x^i}) *_{\theta} \phi. \label{PSSchro}
\end{align} 
%
%%%%%%%%%%%%%%%%%%%%%%%%%%%%%%%%%%%%%%%%%%%%%%%%%%%%%%%%%%%%%%%%%%%%%%
\subsection{Generalized Heisenberg Equation in Phase Space}
%%%%%%%%%%%%%%%%%%%%%%%%%%%%%%%%%%%%%%%%%%%%%%%%%%%%%%%%%%%%%%%%%%%%%%
%
If $H$ is time-independent, then the finite counterpart of infinitesimal change (\ref{Moyal1}) becomes,
\begin{align}
    f(x,p,t) = e_*^{(1/i\hbar) H t} * f(x,p) * e_*^{-(1/i\hbar) H t},\nonumber
\end{align} 
where
\begin{align}
     e_*^{(1/i\hbar) H t} &= 1 + \frac{t}{i\hbar} H  + \frac{1}{2!}\left(\frac{t}{i\hbar}\right)^2  H * H + \ldots \\
     e_*^{(1/i\hbar) H t} * e_*^{-(1/i\hbar) H t} &= e_*^{-(1/i\hbar) H t} * e_*^{(1/i\hbar) H t}=1\nonumber
\end{align} 
Using the property (\ref{cyclic}), the expectation value may be written as
\begin{eqnarray}
\begin{aligned}
    \left< A \right> &= \int  d^n x \, d^n p \, \left(e_*^{-(1/i\hbar) H t} * A(x,p)* e_*^{(1/i\hbar) H t}\right) * f(x,p)  \label{expect2}  \\
     &=\int  d^n x \, d^n p \, {\mathcal A}(x,p,t)* f(x,p).
\end{aligned}
\end{eqnarray} 
where we have defined the function
\begin{align}
     {\mathcal A}(x,p,t) = e_*^{-(1/i\hbar) H t} * A(x,p)* e_*^{(1/i\hbar) H t}. \nonumber
\end{align} 
The time derivative of this function turns out to be
\begin{align}
     \frac{d\, {\mathcal A}}{dt} = \frac{1}{i\hbar}[{\mathcal A} \scomma H] \label{Heisen}
\end{align} 
which is the phase space analog of the Heisenberg equation. 
%
%%%%%%%%%%%%%%%%%%%%%%%%%%%%%%%%%%%%%%%%%%%%%%%%%%%%%%%%%%%%
\subsection{Heisenberg Equation}
%%%%%%%%%%%%%%%%%%%%%%%%%%%%%%%%%%%%%%%%%%%%%%%%%%%%%%%%%%%%
%
From (\ref{expect2}), it follows that
\begin{align}
    \left< A \right> &= \int  d^n x \, d^n p \, \left(e_*^{-(1/i\hbar) H t} * A(x,p)* e_*^{(1/i\hbar) H t}\right) * f(x,p)   \nonumber \\
     &= \int  d^n x \, d^n p \, d^n w \, \left(e_*^{-(1/i\hbar) H t} * A(x,p)* e_*^{(1/i\hbar) H t}\right) * \phi(x)* e^{ip_i w^i} * \phi^*(x) \nonumber \\
     &= \int  d^n x \, d^n p \, d^n w \, \left\{\left(e_{*_{\theta}}^{-(1/i\hbar) {\hat H} t} *_{\theta} A(x,-i\hbar\frac{\partial}{\partial x^i})*_{\theta} e_{*_{\theta}}^{(1/i\hbar) {\hat H} t}\right) * \phi(x)\right\} * \nonumber \\
     & \phantom{\int  d^n x \, d^n p \, d^n w \,} * e^{ip_i w^i} * \phi^*(x) \nonumber \\
     &= \int  d^n x \, d^n p \,d^n w \, \left\{ {\hat {\mathcal A}}(x,-i\hbar\frac{\partial}{\partial x^i}, t) * \phi(x)\right\}* e^{ip_i w^i} * \phi^*(x)\nonumber
\end{align} 
where in the third step, we have made use of (\ref{T2C2}) and defined the operator $ {\hat H} = H(x,-i\hbar\frac{\partial}{\partial x^i})$. And in the last step we have defined the operator
\begin{align}
     {\hat {\mathcal A}}(x,-i\hbar\frac{\partial}{\partial x^i}, t) = e_{*_{\theta}}^{-(1/i\hbar) {\hat H} t} *_{\theta} A(x,-i\hbar\frac{\partial}{\partial x^i})*_{\theta} e_{*_{\theta}}^{(1/i\hbar) {\hat H} t}.\nonumber
\end{align} 
The time derivative of the above operator leads to the Heisenberg equation:
\begin{align}
     \frac{d\, {\hat {\mathcal A}}}{dt} = \frac{1}{i\hbar}[{\hat {\mathcal A}} \scomma H] \label{Heisen1}
\end{align} 
%
%%%%%%%%%%%%%%%%%%%%%%%%%%%%%%%%%%%%%%%%%%%%%%%%%%%%%%%%%%
\section{Twisted Galilean Symmetry}
%%%%%%%%%%%%%%%%%%%%%%%%%%%%%%%%%%%%%%%%%%%%%%%%%%%%%%%%%%
%
An important difference between the commutative and NC WDFs is that the integration over $p$ of (\ref{ncwdf}) gives the quantity $ \phi(x)*_{\theta}\phi^*(x)$ which is nonlocal. In the case $\theta=0$, this marginal distribution is the quantum mechanical probability density \cite{Wigner,Hillery}. In fact, it is one of the requirements, apart from Galilean symmetry and others, that Wigner imposed on $f(x,p)$ to make it a quantum mechanical phase space distribution function. If we extend the same definition to $\theta \neq 0$ case, then the quantum mechanical probability density in NCQM becomes nonlocal (but see \cite{Barbosa} where the local quantity $ \phi(x)\phi^*(x)$ is treated as the probability density in NCQM). 

Another important difference is that (\ref{ncwdf}) is invariant not under Galilean transformation but under 'twisted' Galilean transformation. In conventional QM, WDF (\ref{cwdf}) is Galilean invariant \cite{Wigner,Hillery}. Galilean transformation consists of two parts: (i) if $\phi(x) \to \phi(x+a)$ then $ f_c(x,p) \to f_c(x+a,p)$ and (ii) if $ \phi(x) \to e^{ip'_i x^i} \phi(x)$ then $f_c(x,p) \to f_c(x,p-p')$. The second part may also be stated as follows:  if $ \tilde{\phi}(k) \to \tilde{\phi}(k-(p'/\hbar))$ then $f_c(x,p) \to f_c(x,p-p')$. 

In the case of NCQM, since the translations $x'_i \to x_i + a_i $ preserve (\ref{canonical}) the first part is trivially satisfied by (\ref{ncwdf}).  To deal with the second part, we write (\ref{ncwdf}) using Fourier expansion of the wave function:
\begin{align}
     f(x,p) &= \int \frac{d^n w}{(\pi\hbar)^n} \frac{d^n k}{(2\pi)^n}\,\frac{d^n k'}{(2\pi)^n} \, e^{-\frac{2i}{\hbar} p_i w^i} \tilde{\phi}(k) \tilde{\phi^*}(k') (e^{ik_i(x^i+w^i)} *_{\theta} e^{-ik'_i(x^i-w^i)}). \label{ncwdf3}
\end{align} 
Under twisted Galilean transformation, the product $ \tilde{\phi}(k) \tilde{\phi^*}(k) $ transforms as follows \cite{Chakraborty}:
\begin{align}
     \tilde{\phi}(k) \tilde{\phi^*}(k') \to \tilde{\phi}(k_i - \frac{p'_i}{\hbar}) \tilde{\phi^*}(k'_i -\frac{p'_i}{\hbar}) e^{\frac{i\theta^{ij}}{2} p'_i (k_j-k'_j)}, \label{twistGali}
\end{align} 
where $\vec{p'}$ is a change in the momentum of the particle as a result of a boost along an arbitrary direction. In the case of ordinary Galilean transformation the exponential factor in (\ref{twistGali}) is absent. Using (\ref{twistGali}) in (\ref{ncwdf3}) and performing the change of integration variables $l_i = k_i - (p'_i/\hbar) $ and $l'_i = k'_i - (p'_i/\hbar) $ in (\ref{ncwdf3}), we get 
\begin{align}
     f'(x,p)&= \int \frac{d^n w}{(\pi\hbar)^n} \frac{d^n l}{(2\pi)^n}\,\frac{d^n l'}{(2\pi)^n} \, e^{-\frac{2i}{\hbar} (p_i-p'_i) w^i} \tilde{\phi}(l) \tilde{\phi^*}(l') \times\nonumber \\
     & \phantom{\int \frac{d^n w}{(\pi\hbar)^n}\frac{d^n l}{(2\pi)^n}} \times \left(e^{il_i(x^i+w^i)} *_{\theta} e^{-il'_i(x^i-w^i)}\right) \nonumber \\
     & = f(x,p-p'). \nonumber 
\end{align} 
In the absence of the exponential factor in (\ref{twistGali}), $f'(x,p) \neq f(x,p-p') $ indicating that $f(x,p)$ possesses twisted Galilean symmetry.
%
%%%%%%%%%%%%%%%%%%%%%%%%%%%%%%%%%%%%%%%%%%%%%%%%%%%%%%%%%%%%%
\section{Concluding Remarks}
%%%%%%%%%%%%%%%%%%%%%%%%%%%%%%%%%%%%%%%%%%%%%%%%%%%%%%%%%%%%%
%
We considered the NC phase space in which the positions do not commute among themselves and with canonically conjugate momenta. We started with the probabilistic description of one-particle classical mechanics and introduced the noncommutativity through the concept of deformed $*$-product. We showed that the bare NC geometrical structure of phase space and a normalized phase space distribution function with its quantum condition and with the 'classical' definition of expectation value for a physical observable are enough to extract (i) the whole formalism of noncommutative QM of nonrelativistic spinless particles in phase space with its eigenvalue equation and the Moyal dynamical equation and (ii) the whole formalism of NCQM of the same in Hilbert space including the concept of wave function, the operator algebra, Schr\"{o}dinger equation and the Heisenberg equation. The quantum condition we use is $f*f=\kappa f$ for the phase space distribution function $f(x,p)$, which can be roughly inferred from the square of Dirac delta distribution in the classical case. From the quantum condition we constructed  NC Fredhold integral equation of second kind and showed that NC Hilbert-Schmidt theory of integral equation demands that $1/\kappa$ be directly related to the deformation parameter and that $1/\kappa$ be integrally quantized. We identified that integer with the number of non-interfering equally probable mixed states. For a pure state the solution of that integral equation led to NC Wigner distribution function which we showed to possess twisted Galilean symmetry. By extremizing the expectation value for a physical observable, we derived eigenvalue equations in both the formalisms and identified the point of separation of the two formalisms. We could also elegantly derive the time dependent Schr\"{o}dinger  equation by working out the quantum condition preserving variations of $f$ that would yield the correct dynamical equations in the classical limit.  
 
Finally, we believe that the formulation of NCQM(QM) outlined here might lead to a deeper understanding of quantum formalism viewed from the classical domain, and also that it might serve as a guiding principle to the phase space functional approach to the quantization of Nambu brackets \cite{Nambu}, and to derive the corresponding 'Hilbert space' formalisms of them. We hope to pursue these problems in future. 
%
%%%%%%%%%%%%%%%%%%%%%%%%%%%%%%%%%%%%%%%%%%%%%%%%%%%%%%%%
\acknowledgments
%%%%%%%%%%%%%%%%%%%%%%%%%%%%%%%%%%%%%%%%%%%%%%%%%%%%%%%%
%
It's a pleasure to thank Prof. P. Mitra for a reading of the manuscript and for discussions on it. The useful comments of the referee are also acknowledged. 
%
%%%%%%%%%%%%%%%%%%%%%%%%%%%%%%%%%%%%%%%%%%%%%%%%%%%%%%%%

\end{document}